\PassOptionsToPackage{sort&compress}{natbib}
\PassOptionsToPackage{svgnames}{xcolor}
\documentclass[utf8]{FrontiersinVancouver} 

\setcitestyle{square} 
\usepackage{url,lineno,microtype}
\usepackage[onehalfspacing]{setspace}
\usepackage[T1]{fontenc}
\usepackage[
    pdftitle={SMASH workflow},%
    colorlinks,
    linkcolor=FireBrick,    
    citecolor=ForestGreen,  
    filecolor=magenta,      
    urlcolor=MediumBlue     
]{hyperref}
\usepackage[nameinlink]{cleveref}         

\crefname{section}{Section}{Sections}
\crefname{figure}{Figure}{Figures}
\crefname{table}{Table}{Tables}
\crefname{equation}{Eq.}{Eqs.}

\def\keyFont{\fontsize{8}{11}\helveticabold }
\def\firstAuthorLast{A.~Sciarra and H.~Elfner}
\def\Authors{Alessandro Sciarra\,$^{1}$ and Hannah Elfner\,$^{2,1,3,4,*}$}
\def\Address{$^{1}$Institute for Theoretical Physics\\Goethe University, Frankfurt am Main, Germany \\
$^{2}$GSI Helmholtzzentrum für Schwerionenforschung\\ Darmstadt, Germany\\
$^{3}$Frankfurt Institute for Advanced Studies\\  Frankfurt am Main, Germany\\
$^{4}$Helmholtz Research Academy Hesse (HFHF)\\GSI Helmholtz Center, Campus Frankfurt, Frankfurt
am Main, Germany }
\def\corrAuthor{Hannah Elfner}
\def\corrEmail{h.elfner@gsi.de}

\newcommand{\heavyion}{\mbox{heavy-ion}}
\newcommand{\FAIR}[1]{\textcolor{Navy}{\textit{#1}}}

\begin{document}
\onecolumn
\firstpage{1}

\title[SMASH event generator]{SMASH as an event generator for \heavyion\ collisions}

\author[\firstAuthorLast]{\Authors} 
\address{} 
\correspondance{} 

\extraAuth{}

\maketitle

\begin{abstract}

In this article we present an overview of the SMASH hadronic transport approach that is applied for non-equilibrium dynamics of hadrons in \heavyion\ collisions.
We will give an overview about the ingredients of the approach and the applications for the dynamical description of \heavyion\ collisions and for calculations of fundamental properties of the hadron gas.
The main emphasis of the article will be the infrastructure for sustainable software development that we have developed over the last 10 years including extensive unit tests and continuous integration.
We will also provide one section about the performance of the code and how it can be analyzed and improved in the future.

\tiny
\keyFont{ \section{Keywords:} Transport Theory, Sustainable Software, Performance, Event Generator, Heavy-ion collisions} 
\end{abstract}

\section{Introduction}

The building blocks of matter and the fundamental forces of nature are summarized in the standard model of particle physics.
Besides the electromagnetic and the weak interaction, the strong force is responsible for keeping quarks together.
Gluons are the corresponding exchange particles and bind the quarks in pairs and triples such that color neutral objects are formed, the hadrons.
Studying hadronic matter under extreme conditions of temperature and density is the main goal of \heavyion\ research.
Within the collisions of nuclei that are accelerated to almost the speed of light by major accelerators around the world~---~currently the Large Hadron Collider (LHC) at CERN, the Relativistic Heavy Ion Collider (RHIC) at BNL and the SIS-18 at GSI~---~the matter is heated and compressed such that the circumstances are similar to the ones only microseconds after the Big Bang or the ones in neutron star mergers \cite{LIGOScientific:2017vwq}.

At such high temperatures and/or densities, hadronic matter undergoes a phase transition to the \mbox{quark-gluon} plasma, where hadrons are not bound anymore and quarks and gluons become the active degrees of freedom. To study the properties of this new state of matter as well as the nature of the transition are the main motivations behind \heavyion\ research.
Experimentally, it is only possible to observe the final state particle distributions of hadrons and electromagnetic probes in the detector.
To connect the measurements to fundamental insights, detailed models of the dynamical evolution are indispensable.
The hadronic transport approach SMASH (Simulating Many Accelerated Strongly-interacting Hadrons) \cite{SMASH:2016zqf,wergieluk_2024_10707746} is one of the general purpose event generators in the field.
This microscopic transport approach incorporates the non-equilibrium dynamics of hadrons in a \heavyion\ collision and is therefore applicable for the full evolution at low beam energies and for the early/late stages at high beam energies.

Transport approaches have been developed and applied for more than 30 years and many approaches exist \cite{Hartnack:1989sd,Bass:1998ca,Nara:1999dz,Lin:2004en,Bratkovskaya:2011wp,Buss:2011mx,Pierog:2013ria,Glassel:2021rod} that are available to the public, at least on request.
It is crucial to compare different approximations and frameworks carefully to assess the impact of certain choices on experimental observables.
This has been carried out extensively by the Transport Model Evaluation Project \cite{TMEP:2017mex,TMEP:2019yci,TMEP:2022xjg}. This exemplifies why open source codes and transparent development practices are important to allow the user to assess the validity of the approach and understand changes over time, when new features are included.
Another aspect is that a modern software development infrastructure allows new students and external collaborators \cite{Sorensen:2020ygf, Sorensen:2021zme, Oliinychenko:2022uvy} to access and work with the code without much overhead.

In this work, the main ingredients and scope of the SMASH transport approach are explained in \cref{sec:physics}.
In \cref{sec:software} the software development environment is described in detail to show-case an example of transparent development in line with the FAIR4RS rules.
\Cref{sec:perf} addresses performance and optimization potential of the code, since it is also part of sustainable software development to use high performance computing resources efficiently.
The last \cref{sec:sum} summarizes the main findings and conclusions.

\section{Physics Application}
\label{sec:physics}

SMASH is a relativistic hadronic transport code, that represents an effective solution of the relativistic Boltzmann equation incorporating all well-known hadronic species based on the particle data book from 2018 \cite{ParticleDataGroup:2018ovx}

\begin{equation}
\label{eq:boltzmann_equation}
  p^\mu \partial_\mu f_i(x,p) + m_i F^{\alpha} \partial^{p}_{\alpha} f_i(x, p) = C^i_{\rm coll}
\end{equation}
where $f_i$ is the distribution function, $C^i_{\rm coll}$ is the collision term, $F^{\alpha}$ is the force experienced by individual particles and $m_i$ is the particle mass.
For high beam energy collisions, the potential term can be neglected $F^{\alpha} = 0$, while for low beam energy collisions, $F^{\alpha} = -\partial^{\alpha} U(x)$ where $U(x)$ is the mean-field potential.
The collision term contains all the details of the interactions between the particles.
At lower energies the reactions proceed via resonance excitation and decay while at high energies strings are excited and fragment via Pythia \cite{Bierlich:2022pfr,10.21468/SciPostPhysCodeb.8-r8.3}.
Within SMASH binary collisions are handled via a geometric collision criterion, but there is also the stochastic rates approach available relying on testparticles and interaction probabilities on a spatial grid, which allows for multi-particle reactions.

\begin{figure*}
    \centering
    \includegraphics[width=\textwidth]{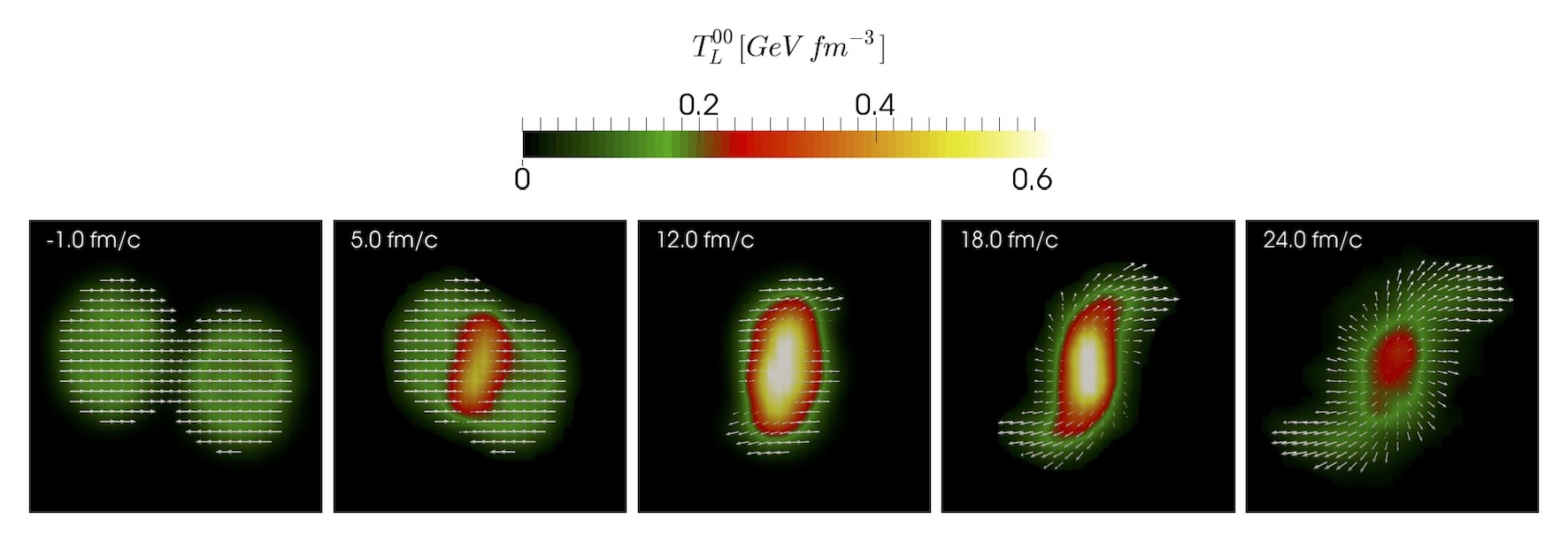}
    \caption{
      Figure taken from \cite{SMASH:2016zqf}.
      Snapshots of the time evolution of a  Au+Au collision at $E_\text{kin} = 0.8A\,\text{GeV}$ with impact parameter $b = 3\,\text{fm}$, $N_\text{test} = 20$.
      The energy density $T^{00}_L$ (background color) and velocity of Landau frame (arrows) are displayed for the baryons in the system.
      The velocity is proportional to the arrow length and the maximal arrow length corresponds to a velocity of 0.55 $c$.
    }
    \label{fig:snapshots}
\end{figure*}

\Cref{fig:snapshots} displays several steps of the time evolution for a low energy heavy ion collision corresponding to the ones studied at SIS-18 at GSI.
One can clearly see the hot and dense region that is formed as well as the collective velocity of the system.
The collective flow is one of the major observables to constrain transport coefficients and the equation of state of strongly interacting matter.

\begin{figure*}
    \centering
    \includegraphics[width=0.92\textwidth]{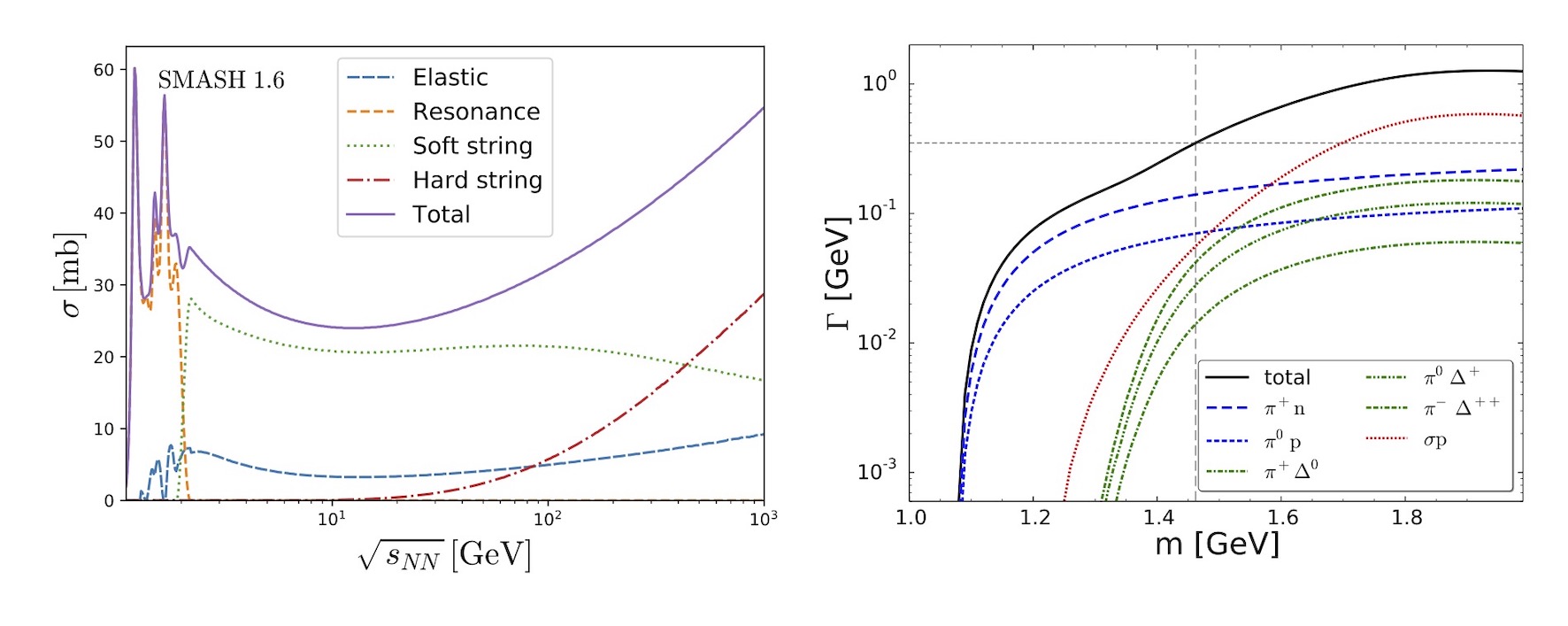}
    \caption{
      Figures taken from \cite{SMASH:2016zqf, Mohs:2019iee}.
      Left: Cross-section for pp reactions as a function of the center-of-mass energy, the contributions of different processes are shown.
      Right: Mass-dependent width of the $N^*(1440)$ resonance indicating that different branching ratios appear at different thresholds.
    }
    \label{fig:interactions}
\end{figure*}

\Cref{fig:interactions} (left) shows as an example the energy dependence of the proton-proton cross-sections.
At low energies, resonance excitations dominate, followed by a regime of soft string excitation and decay while hard processes take over at high energies.
The elastic cross-section is fitted to the available experimental data.
In \cref{fig:interactions} (right) the mass dependent width of the $N^* (1440)$ resonance is shown.
The width is inversely proportional to the lifetime of the resonance.
The mass of the resonance is determined by a relativistic Breit-Wigner distribution in each individual scattering.
Dependent on the mass, different decay processes are kinematically possible, therefore the width is changing accordingly.
These figures are shown to give an impression of the complexity of modeling all the individual reaction channels of more than 150 hadronic species.

SMASH can be applied in several settings; \texttt{Collider} for full collision simulations as shown in \cref{fig:snapshots}, \texttt{Box} for infinite hadronic matter, \texttt{Sphere} for expanding matter as well as \texttt{List} where the user can insert any particle list of choice to be evolved in time.
Examples of calculations in infinite hadronic matter include the calculation of transport coefficients of the hadron gas at different temperatures and chemical potentials (see e.g. \cite{Rose:2017bjz, Rose:2020sjv}).
The expanding sphere serves as a simplified setup to simulate collision dynamics as has been employed e.g. in \cite{Dorau:2019ozd}.
The listmodus is crucial for calculations where SMASH is employed for the late hadronic rescattering as part of hybrid approaches.
The standard model for the dynamical evolution of high energy \heavyion\ collisions involves a non-equilibrium initial state, a viscous hydrodynamic evolution and hadronic rescattering. SMASH has been applied for this purpose in our own SMASH-vHLLE hybrid\footnote{\url{https://github.com/smash-transport/smash-vhlle-hybrid}} approach \cite{Schafer:2021csj,Gotz:2022naz} as well as by other theory groups around the world  \cite{Nijs:2020roc, JETSCAPE:2020mzn, JETSCAPE:2020shq,Wu:2021fjf}.
It is also possible to couple to SMASH as a library to employ certain modules or just the hadron gas properties.
The complete particle content and the resonance properties can be changed by exchanging two human readable text file, \texttt{particles.txt} and \texttt{decaymodes.txt}, which offers a lot of flexibility.

An additional target group of the approach are large experimental collaborations like ALICE, STAR, NA61, CBM and HADES that are running the code to make predictions or compare to their measurements.
To ensure that external users can follow updates of the codes and how changes and addition of features affect the physics observables a physics analysis suite is run for each tagged version of the code.
The results for each SMASH version are publicly available\footnote{\url{http://theory.gsi.de/~smash/analysis_suite/SMASH-3.1/index.html}} and the corresponding analysis code is released\footnote{\url{https://github.com/smash-transport/smash-analysis}} as well.
This test includes inherent tests such as the comparison of collision rates to analytic expectations and detailed balance tests as well as comparisons to experimental data in vacuum (elementary cross-sections and angular distributions) and hot and dense medium (AA collisions at different beam energies).
To ensure simple application, the installation of the code is documented in detail and docker/singularity images are provided for each tagged public version.
There are also all commonly used output formats supported by SMASH including several stages of OSCAR formats in text and binary versions as well as a ROOT interface.
In addition, there is an interface for HepMC output to connect to the RIVET analysis software that is being adopted by the \heavyion\ community recently \cite{Bierlich:2024vqo}.

\section{Sustainable Software Development}
\label{sec:software}

SMASH is an open source code developed by one working group, and employed by many theoretical groups and experimental collaborations world-wide.
To make such a code sustainable, a few points are important.
The software development follows strict rules that ensure the FAIR4RS principles for software development~\cite{FAIR_software} and, hence, sustainability.
The released code is publicly available on GitHub and published on Zenodo, where each version gets a DOI associated and the \texttt{CITATION.cff} file maintained in the codebase is used for metadata (\FAIR{Findable} and \FAIR{Accessible}).
Internal development of new features and constant improvements happen in a private repository, that is regularly made public, whenever a new version of the software is released (1-2 times per year).
For flawless and straightforward usage, every new release is shipped with Docker images that allow any user to immediately run simulations in a dedicated container.
Versioning is crucial to avoid misunderstandings and misinterpretation of physics results.
Typically, in software development, semantic versioning is used, where the first digit highlights major changes, the second digit indicates backward compatible changes and the third digits is reserved for small bug fixes.
SMASH does not fully comply with this versioning scheme, as new physics features are often driving new releases and backward compatibility is not always easy to guarantee.
However, a detailed \texttt{CHANGELOG} file is kept up-to-date during development and any potentially relevant change for the user is listed and shortly described there.
Every new release is provided with both an extensive code documentation as well as a less technical user guide (\FAIR{Reusable}).

\begin{figure*}
    \centering
    \includegraphics[width=0.45\textwidth]{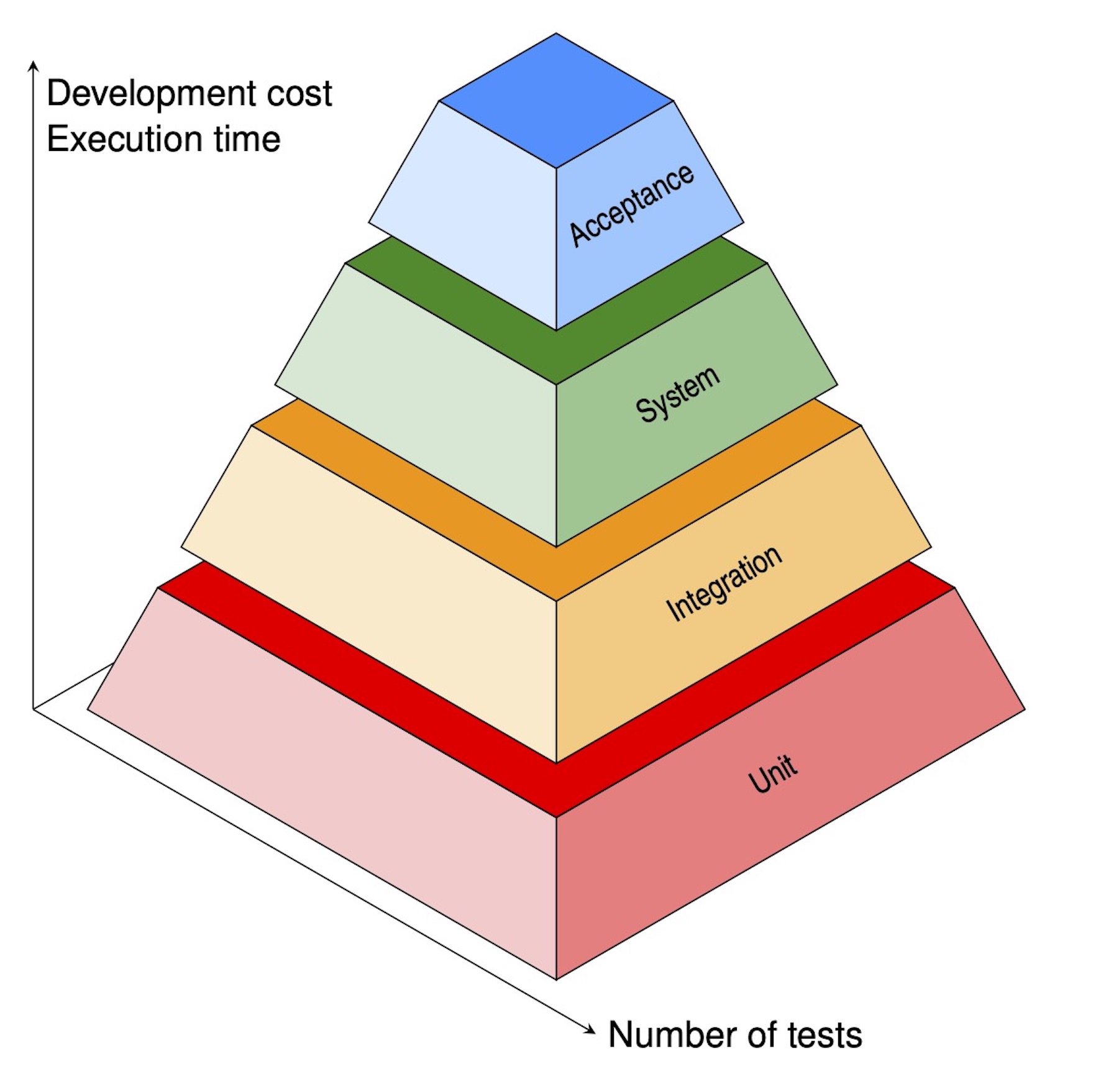}
    \caption{
        The tests pyramid, taken from~\cite{Clean_testing}.
        Different types of software tests are classified according to some features.
        The base surface of each layer is proportional to how many of that test type should exist in a codebase and it is inversely proportional to their execution time and development cost.
        Roughly speaking, unit tests are about testing each unit of the codebase.
        A unit might be e.g.\ a module or sometimes a class/function.
        Integration tests are meant to test the correct interaction between different units.
        System tests ensure the correctness of larger (if not all) parts of a codebase.
        Acceptance tests determine whether the application meets the requirements.
        In our context, they ensure that physics is correctly described and e.g.\ experimental data reproduced as expected.
    }
    \label{fig:tests}
\end{figure*}

No large codebase could survive, grow and improve over many years without a rigorous and complete testing strategy or if good practices in software development were not applied as often as possible.
Unit, integration and system tests are regularly developed in SMASH to ensure code correctness and over 100 test suites exist.
\Cref{fig:tests} illustrates how software tests are classified and what is usual to be expected in terms of execution time and number of tests.
SMASH acceptance tests are developed in \href{https://github.com/smash-transport/smash-analysis}{a separate codebase} and consists of a whole plethora of physics tests already mentioned in \cref{sec:physics}.

New developments happen on dedicated Git branches and are merged into the private codebase via pull requests, a very well known GitHub functionality, which offers the possibility to associate given requirements before authorizing the merge.
These are a positive code review and the successful run of a given set of actions.
The former is thoughtfully done by at least another group member, although often many developers interact exchanging suggestions and improvements.
The latter, instead, is an automatic continuous integration system which is set up to build the code on a few common architectures and with different compilers, to ensure proper formatting as well as complete documentation of the code and to check that all unit, functional and system tests pass.
If any of these operations fail, the merge of the pull request is denied.
Due to their intense computational cost, acceptance tests cannot be run at each pull request, nor would it make sense.
A natural point in development to run them is just before each new public release, in order to ensure transparency between versions.

Given the available tests infrastructure as well as the procedure to merge code changes into the codebase, developers are free to experiment and refactor code.
This is a key aspect in order to be able to constantly apply clean code~\cite{martin2009clean} principles and, hence, keep the software development sustainable also from the technical point of view of changing existing code and writing new one.

The software is written in C++ and built via CMake and is modular, such that it is usable in parts or fully as a library (\FAIR{Interoperable}).
Usage of new C++ standards is constantly considered, although the need to offer a reliable product on most machines in the world delays upgrades to most recent language features (at the moment C++17 is used).

\section{Performance and Optimization}
\label{sec:perf}

SMASH performance is a core aspect in development and there is lots of ongoing effort trying to find possible improvements.
However, \textit{«Programmers waste enormous amounts of time thinking about, or worrying about, the speed of noncritical parts of their programs, and these attempts at efficiency actually have a strong negative impact when debugging and maintenance are considered. We should forget about small efficiencies, say about 97\% of the time: premature optimization is the root of all evil. Yet we should not pass up our opportunities in that critical 3\%.»}~--~D.~Knuth~\cite{Knuth_1974}.
This is the reason why optimization tasks are kept as much as possible separated from development and clean-code refactoring.
Roughly speaking the code has first to work~---~tests ensure that~---~and then, exactly because tests exist, the code can be freely changed looking for better performance.
This is only done after having thoroughly profiled the code, typically measuring how many times single functions are called and how much of SMASH runtime is spent executing each of them.
In this way, educated guesses about possible bottlenecks can be made and optimization efforts structured.

\begin{figure*}
    \centering
    \includegraphics[width=\textwidth]{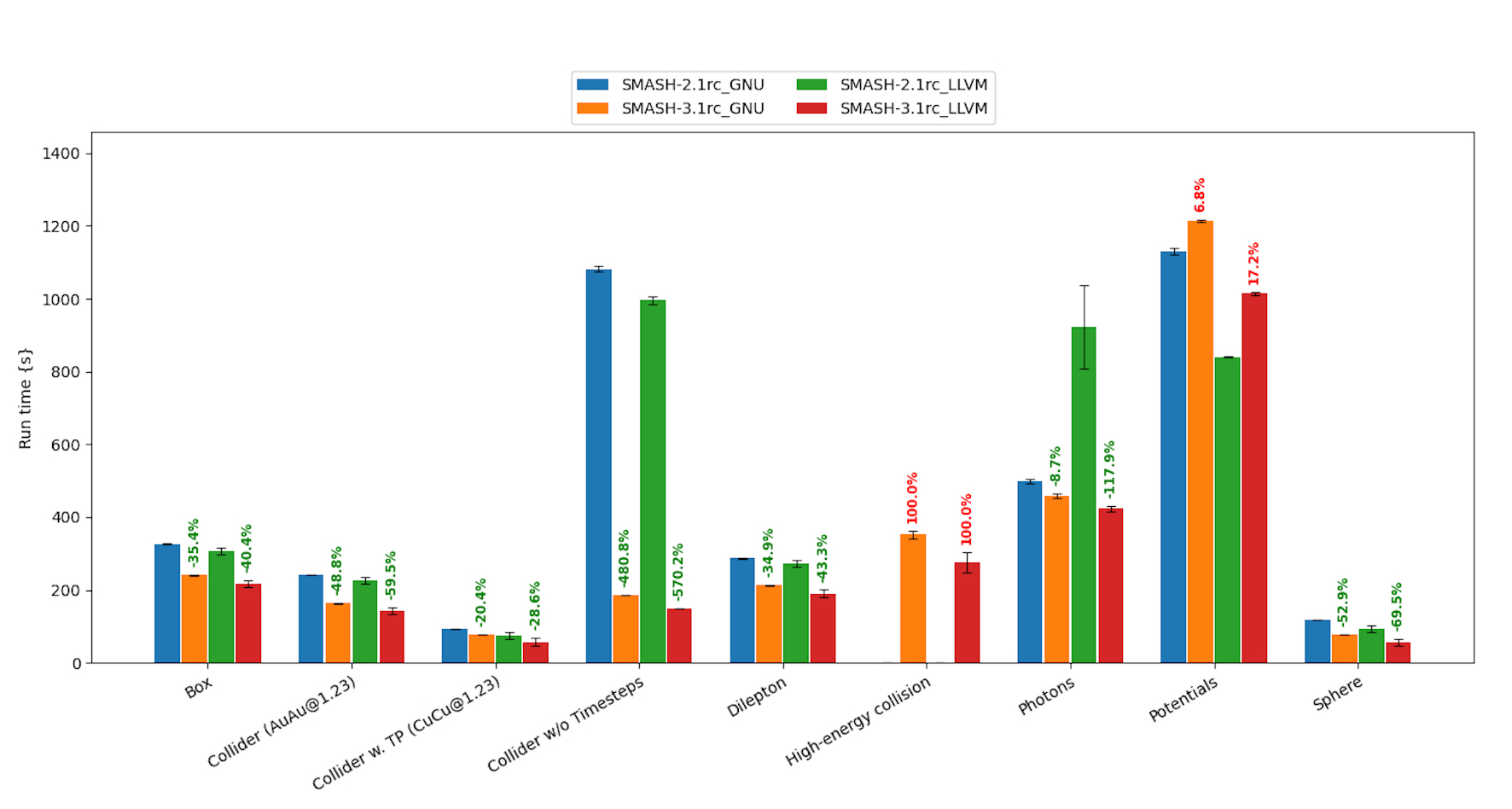}
    \caption{
        Evolution of SMASH performance over 4 releases.
        Different groups of bars refer to different SMASH setups.
        The percentage above the second and fourth bar within each group indicates the relative improvement (green) or loss (red) of performance.
        Since the ``high-energy collision benchmark'' has been introduced in SMASH-2.2, two bars are missing there.
        The potentials implementation is known to be sub-optimal and performance has slightly degraded there, because of some code-entanglement which will be cleaned up soon.
        \texttt{Box}, \texttt{Collider} and \texttt{Sphere} refer to the SMASH default setups already mentioned in \cref{sec:physics}, while the other names refer to different physics systems and/or observables, possibly simulated with different approaches.
        The \textbf{\texttt{bin/benchmarks}} folder in SMASH codebase contains all further details.
    }
    \label{fig:benchmark}
\end{figure*}

In a complex codebase, no matter how carefully development is done, accidental pessimizations can occur during development.
This is why SMASH is shipped with a standardized set of benchmarks which are run on the same machine before every new release, in order to check how the new version performs with respect to the previous one(s).
As different compilers produce different executables in terms of performance, benchmarks are run compiling the code using two of the most common C++ compilers: GNU and LLVM.
In \cref{fig:benchmark} the benchmark outcome comparing two given SMASH versions is visualized.
Overall the trend is clear: The codebase performance has constantly improved and in the last couple of years SMASH got roughly two times faster in most of the possible setups.

\section{Summary and Conclusions}
\label{sec:sum}

SMASH is a general purpose event generator for \heavyion\ collisions based on the relativistic Boltzmann equation.
The microscopic non-equilibrium dynamics of all hadrons are simulated from initial to final stage or it is employed for the late hadronic rescatterings at high beam energies.
The way how sustainable software development is employed is pioneering in the landscape of research software in the field of theoretical \heavyion\ reactions.
Therefore, the steps that are undertaken to ensure transparency and reproducibility of results are described in detail.
The initial performance analysis and first optimization have been shown as well, but there is certainly room for additional work in the future.
For example, one can look into multi-threading or parallel computing for CPU intensive settings involving stochastic rates, many testparticles or parallel ensembles.

\section*{Conflict of Interest Statement}

The authors declare that the research was conducted in the absence of any commercial or financial relationships that could be construed as a potential conflict of interest.

\section*{Author Contributions}

H.E. is the maintainer of the SMASH code and has written the sections about the physics application, introduction and summary and has contributed to the other sections.
A.S. has performed the optimization studies as well as major refactoring in the last years.
He constantly works to keep the codebase as clean and high-quality as possible and has written the sections on sustainable software development and about the code performance.

\section*{Funding}
Funded by the Deutsche Forschungsgemeinschaft (DFG, German Research Foundation) – Project number 315477589 – TRR 211.

\section*{Acknowledgments}
Computational resources have been provided by the Center for Scientific Computing (CSC) at the Goethe-University of Frankfurt and the GreenCube at GSI.

\bibliography{SMASH-as-event-generator-for-HIC}


\end{document}